\documentclass[12pt,spanish,twosides]{article} 
\usepackage{hyperref}
\pdfoutput=1
\providecommand{\keywords}[1]{\textbf{\textit{keywords:}} #1}

\newcommand{\proof}{\noindent {\bf Proof. }}

\newcommand{\qed}{\hfill \fbox{} \vspace{.3cm}}

\newtheorem{definition}{Definition}

\newtheorem{proposition}{Proposition}

\begin{document}

\title{On the Integrability of the Geodesic Flow on a Friedmann-Robertson-Walker Spacetime}
\vspace{1cm}
\author{Francisco Astorga\footnote{fastorga@ifm.umich.mx}, J. Felix Salazar\footnote{jfelixsalazar@ifm.umich.mx} and  Thomas Zannias\footnote{zannias@ifm.umich.mx} \\\
\\
\footnotesize{Instituto de F\'isica y Matem\'aticas, Universidad Michoacana de San Nicol\'as de Hidalgo,}\\
\footnotesize{Edificio C-3 , Ciudad Universitaria, 58040 Morelia, Michoac\'an, M\'exico.}\\
}

\maketitle

\begin{abstract}
We study  the geodesic flow on the 
cotangent bundle $T^{*}M$ of a 
Friedmann-Robertson - Walker (FRW) spacetime $(M,g)$.
 On this bundle, the
 Hamilton-Jacobi equation
 is completely separable
 and this property allows us  to 
 construct four linearly independent integrals in involution, i.e.
 Poisson commuting
 amongst themselves and pointwise linearly independent. 
 As a consequence, the geodesic flow on an FRW background
is completely integrable in the Liouville-Arnold sense. 
For  a spatially flat or spatially closed universe, we 
construct 
submanifolds 
that remain invariant under the action of the flow.
For a spatially closed universe these submanifolds are
topologically
$R\times S^{1}\times S^1\times S^1$, while
for a spatially flat universe they are topologically
$R\times R\times S^1\times S^1$. However, due to the highly symmetrical 
nature of the
background
spacetime, the four integrals in involution 
also admit
regions where they fail to be linearly independent. We identify these regions
although we have not 
been able in a mathematically rigorous fashion
to describe the structure of the associated
invariant submanifolds. Nevertheless,
the phase space trajectories contained
 in these submanifolds  when projected
on the base manifold
describe radial timelike geodesics 
or timelike geodesics  
''comoving'' with
the cosmological expansion.

\end{abstract}


\vskip0.1cm
\section*{Introduction}

In the past, studies of the geodesic flow 
associated with a Riemannian metric
belonged to a field explored  exclusively by 
mathematicians who aimed to analyze the global behavior of 
 geodesics 
on a given Riemannian manifold
(for an overview the reader is refer to: \cite{A},\cite{P},\cite{0},\cite{MO1},\cite{FO}).
As it turns out, 
geodesics are affected by 
the topological and geometrical properties
of the background
manifold
and disentangling 
these effects 
 constitutes 
 a problem of 
immense mathematical complexity. A systematic treatment of the geodesic flow
begins by passing  to an equivalent 
 Hamiltonian system
defined on 
the cotangent bundle\footnote{ 
For an introduction to this bundle and an introduction to the symplectic geometry see for instance
\cite{1},\cite{2},\cite{3}.}
of the underlying manifold
and in that manner
the flow defined by the corresponding
Hamiltonian vector field is what mathematicians refer to as 
the geodesic flow. The ''phase portrait'' of the resulting dynamical system
provides insights on the global behavior of the geodesics on the background manifold
and this ''phase space'' description of the problem
 has been proven to be a fruitful one\footnote{It is worth mentioning
 that an alternative treatment of issues related to the global behavior of geodesics
is based on  Lie point symmetries, i.e.
continuous Lie groups of transformations acting on a suitable space
leaving the geodesic equation form invariant (for an introduction
see for instance refs.\cite{O},\cite{B},\cite{S}).
Determining the Lie group of point symmetries of the
geodesic equation is a major undertaking and these groups has been obtained only for
particular background spacetimes (see for example ref. \cite{LieSym} and references therein).
As far as we are aware, no 
work has been done in 
obtaining 
solutions of the geodesic equation from the knowledge 
of the
Lie group of point symmetries. Although
the Lie point symmetries approach essentially is formulated on the tangent bundle, 
it would be of interest to investigate a possible connection
of this  approach and the Hamiltonian methods
advocated in the symplectic approaches.}.
It allows methods of the symplectic geometry
to be called upon and 
successfully 
addresses
 thorny issues such as
whether the geodesics exhibit a regular or chaotic behavior
or whether a given geometry admits closed geodesics, amongst others.
This field is an area of intense investigations
and for an overview, progress  and open problems
regarding the geodesic flow the reader is referred to the refs: \cite{A},\cite{P},\cite{0},\cite{MO1},\cite{FO},\cite{1},\cite{2}.\\

Geodesic flows defined by Lorentzian metrics have 
 become lately
 a relevant topic
 for relativists. The seminal work by Carter \cite{4} on the 
separability of the Hamilton-Jacobi equation on a 
 Kerr background 
 showed that 
causal  geodesics reveal many properties
of the family of Kerr spacetimes.
For some astrophysical implications 
of the geodesic flow on 
a Kerr background see for instance 
the interesting account
of ref.\cite{Shm1}. 
However, geodesic flows  are relevant in other context as well.
The description of black hole shadows (see for instance \cite{EHT})
employs
congruences of null geodesics
on the underlying black hole background while (cosmological) 
gravitational lensing deals with congruences
of null geodesics on a (perturbed)  Friedmann-Robertson Walker (FRW) cosmology
(for an introduction see \cite{GL} and further references therein). Moreover,
the propagation of the elusive cosmic rays
involves the behavior of timelike geodesics either on a cosmological FRW background or on the 
gravitational field of the Milky way (see for instance \cite{PCRG},
\cite{PCRG1},\cite{PCRG2}).
Another setting where geodesic flows
are relevant is offered by 
 the kinetic theory of relativistic gases. In this theory, it is postulated that  
gas particles between collisions move along future directed timelike geodesics 
of the background metric at least for a gas 
composed of neutral particles.
Therefore, knowledge of the geodesic flow
offers insights into the behavior of such gases.
For an introduction to the  relativistic kinetic theory
 see  refs. \cite{5},\cite{6}
and for  more recent accounts see \cite{O1},\cite{O2},\cite{O3}, \cite{OP1},\cite{OP2}. 
Clearly, in these scenarios  we are 
 not any longer dealing with the behavior of 
a single 
geodesic\footnote
{It should be mentioned that, within the context 
of General Relativity, even the behavior of a single geodesic
is very relevant. It suffices to
recall that the first classical tests of
General Relativity are based on the behavior of geodesics within our solar system.} but rather the focus is on 
the global behavior
of congruences of causal geodesics
and here
geodesic flows are becoming relevant.\\

In this work, we  study the geodesic flow defined
by the family of 
Friedmann-Robertson-Walker  (FRW) spacetimes $(M,g)$.
Motivations for such undertaking
come from 
two independent reasons. From the physical view point, 
future directed null or timelike goedesics
are very important in the cosmological context.
For instance, within the geometric optics approximation
null geodesics 
are the messengers of information regarding properties of remote  cosmological systems
while timelike geodesics describe massive particles such as cosmic rays within 
a cosmological context. The reader is referred to standard textbooks for
cosmological applications of such geodesics (see for example
\cite{COS1},\cite{COS2}).
From the mathematical viewpoint,
as we shall show in this paper, 
the geodesic flow on a 
FRW spacetimes is
a completely soluble model and thus becomes a theoretical laboratory
for analyzing the
complex behavior of completely integrable relativistic geodesic  flows.\\

We begin by introducing  the cotangent bundle 
$T^{*}M$ associated with a background manifold 
$(M,g)$ and for completeness  
we provide a brief description 
of  the natural symplectic structure of this bundle
and some of its basic properties that would be relevant later on. 
We introduce 
the Hamiltonian $H$  whose associated
 Hamiltonian vector
field $L_{H}$  defines the geodesic flow over $T^{*}M$. Because
the background metric has Lorentzian signature, the projection of the flow onto
the base manifold $(M,g)$
describes families of timelike, null or spacelike geodesics. 
For reasons that will become clear further ahead, we
restrict our attention to the
timelike component of this 
flow and our primary focus 
is to investigate
whether this timelike component
is completely integrable in the Liouville-Arnold sense\footnote{For an introduction
to integrable Hamiltonian systems consult refs. \cite{1},\cite{2},\cite{3}.}.
We show that for any Killing vector field
$\xi$ admitted by the background metric $g$, there corresponds an
integral of motion defined over $T^{*}M$. Since any FRW metric admits six linearly independent 
 Killing vector fields,
we construct six integrals whose Poisson bracket with the Hamiltonian $H$
vanishes over 
$T^{*}M$. We find 
that these six integrals fail to be in involution,
 i.e. the Poisson brackets amongst themselves fails to be vanishing.
However, based on the separability of the 
Hamilton-Jacobi equation on an FRW background, we construct 
four new integrals $F_{i},i=(1,2,3,4)$ that are Poisson commuting amongst themselves
and moreover are pointwise linearly independent over
regions of $T^{*}M$ and this establishes
that  the geodesic flow on an FRW is indeed 
completely integrable in the Liouville-Arnold sense.
Although this conclusion is welcomed, unfortunately by itself it does not yield
insights regarding the global behavior of the flow.
By studying
the structure of the exterior product $ dF_{1}\wedge\ dF_{2}\wedge dF_{3}\wedge \ dF_{4}$,
we show that
 for  a
spatially closed FRW universe,
there
exist 
families of  
 $R\times S^{1}\times S^1\times S^1$ submanifolds
 of  $T^{*}M$ 
 that are 
invariant under the flow while 
for  a spatially flat FRW universe
we find invariant submanifolds 
that are topologically
$R\times R\times S^1\times S^1$. 
However that is not the end of the story.
We show 
that the product
$ dF_{1}\wedge\ dF_{2}\wedge dF_{3}\wedge \ dF_{4}$,
 vanishes
over particular regions
of $T^{*}M$ 
and over such regions 
the integrals $F_{i},i=1,2,3,4,$
become linearly dependent.
The nature of the invariant submanifolds over regions where
the integrals in involution
of a completely integrable system 
become dependent, is a very subtle problem and has been the subject of thorough mathematical 
investigations (see for instance refs. \cite{FO1}, \cite{PE1}).
For our part in this paper, we discuss the
role of the background Killing fields  on
the vanishing property of 
 the product $ dF_{1}\wedge\ dF_{2}\wedge dF_{3}\wedge dF_{4}$
 and  we argue-although not in very rigorous mathematical manner-
 that the
singular invariant submanifolds
associated with these regions 
are of lower 
than four dimensions. Moreover
the phase space trajectories 
included in this submanifolds  when projected
on the base manifold
describe either radial geodesics 
or geodesics 
comoving with
the cosmological expansion.\\

The structure of the present paper is as follows: In the next section
we introduce the cotangent bundle
$T^{*}M$ over a space-time $(M,g)$
and discuss some of its basic properties.
We  introduce the family of the Hamiltonian vector fields,
the Lie algebra of observables and define the notion of Liouville-Arnold integrabilty.
In section $2$, we restrict
the spacetime $(M,g)$ to be an FRW spacetime and based on the Killing symmetries of the 
background metric $g$, we construct and study properties of the integrals of motion
associated to  the geodesic flow. In section $3$, we discuss properties of the 
Hamilton-Jacobi equation on an FRW background and this analysis allows to conclude the Liouville-Arnold
integrability property of the geodesic flow. In section
$4$, we introduce and discuss properties of the invariant, by the flow, submanifolds
 and in the conclusion section, we discuss some applications
and open problems.


\section{On the symplectic structure of the cotangent bundle}

In this section, we introduce  the cotangent bundle and some basic tools of  the Hamiltonian dynamics. 
Although the material  is standard,  it has been included partially to set up notation
and partially to introduce
some structures 
that are of crucial importance for the development of this work.\\

The
 cotangent bundle  $T^{*}M$ associated with any  smooth\footnote{In this work, all manifolds involved are assumed to be $C^{\infty}$.
 Whenever other fields are employed, 
 they are assumed to be smooth enough so that
 any operation of differentiation performed upon them is to be  well defined. We should mention that
 even though we begin with a spacetime $(M,g)$, actually for the most part of this section
 the metric $g$ does not play any role. It becomes important in defining the
 Hamiltonian $H$ and the Liouville vector field $L_{H}$
 at the very end of this section.}  
$n$-dimensional spacetime $(M,g)$  is defined by

\begin{equation}
T^{*}M=\{(x,p), x\in M, p\in T^{*}_{x}M \} 
\label{eq1}
\end{equation}
and this 
$T^{*}M$ defines the natural projection map  

\begin{equation}
 \pi:T^*M \mapsto M: (x,p)\mapsto \pi(x,p)=x
 \label{pro}
 \end{equation}
so that 
at any $x\in M$, the 
 fiber $\pi^{-1}(x)$  is isomorphic to the cotangent space $T^{*}_{x}M$.
Moreover the  base manifold $(M,g)$ induces 
 upon $T^*M$ an atlas 
so that $T^*M$ becomes a $2n$-dimensional smooth, orientable manifold.
This can be seeing by noting
that any 
 local chart $ (U,\phi) $ in the $C^{\infty}$ atlas of $(M,g)$
defines the map:
$$
\Psi \!:V=\pi^{-1}(U)\!\to \phi(U)\times R ^{n}: (x,p)\mapsto \Psi(x, p):=
$$
$$
= \phi(\pi(x,p)), p_{x}( \left. \frac {\partial}{\partial x^{1}} \!\right|_{x}\!), p_{x}\left.(\frac {\partial}{\partial x^{2}}\!\right|_{x}\!).....p_{x}(\left.\frac {\partial}{\partial x^{n}} \!\right|_{x}\!)=
$$
 \begin{equation}\label{eq2}
=(x^{1},x^{2},.....,x^{n}, p_{1}, p_{2},.......,p_{n})
\end{equation}
which serves
as a local coordinate system\footnote{The set
$(x^{\mu},p_{\mu}):=(x^{1},x^{2},.......,x^{n}, p_{1}, p_{2},.........,p_{n})$ are the local coordinates 
assigned by $(V,\Psi)$ to the point $(x,p)$ on $V:=\pi^{-1}(U)$. The resulting chart  $(V, \Psi)$ is referred as an
adapted chart and the 
 associated coordinates  are often referred as adapted
coordinates.} over $V:=\pi^{-1}(U)$. 
The family of the charts $\{ (U_\alpha,\phi_\alpha)\}$ in the  $C^\infty$ atlas of $(M,g)$ 
defines a
collection of local charts 
$\{(V_\alpha, \Psi_{\alpha})\}$
 on $T^*M$ 
which constitute a $C^\infty$ atlas of $T^*M$.
Furthermore, it can be checked that 
for any two intersecting 
charts 
$(V_\alpha, \Psi_{\alpha})$ and  $(V_\beta, \Psi_{\beta})$
the Jacobian matrix has positive definite determinant and thus the 
collection $\{(V_\alpha, \Psi_{\alpha})\}$
defines an oriented atlas over $T^*M$.\\

Using this atlas, 
at any $(x,p)\in T^*M$,
we construct the tangent space
$T_{(x,p)}(T^{*}M)$, the cotangent space $T^{*}_{(x,p)}(T^{*}M)$ 
and the tensor algebra 
in the usual manner.
The local coordinates $(x^{\mu},p_{\mu})$ 
generate 
at any $(x,p)$ 
the coordinate basis for $T_{(x,p)}(T^{*}M)$ and for $T^{*}_{(x,p)}(T^{*}M)$ 
described by
\begin{displaymath}
\left\{ \left. \frac{\partial}{\partial x^1} \right|_{(x,p)},
\left. \frac{\partial}{\partial x^2} \right|_{(x,p)},\ldots,
\left. \frac{\partial}{\partial x^n} \right|_{(x,p)},
\left. \frac{\partial}{\partial p_1} \right|_{(x,p)},
\left. \frac{\partial}{\partial p_2} \right|_{(x,p)},\ldots,
\left. \frac{\partial}{\partial p_n} \right|_{(x,p)} \right\}
\end{displaymath}
\begin{displaymath}
\left\{ dx^1_{(x,p)},\ldots,dx^n_{(x,p)},
{dp_1}_{(x,p)},\ldots,{dp_n}_{(x,p)} \right\},
\end{displaymath}
and in terms of these bases, any
 $Z\in T_{(x,p)}(T^{*}M)$ can  be expanded according to:
\begin{displaymath}
Z = X^\mu\left. \frac{\partial}{\partial x^\mu} \right|_{(x,p)}
 + P_\mu\left. \frac{\partial}{\partial p_\mu} \right|_{(x,p)},\qquad
X^\mu = dx^\mu_{(x,p)}(Z),\quad P_\mu = {dp_\mu}_{(x,p)}(Z)
\end{displaymath}
with a similar expansion for the elements of $T^{*}_{(x,p)}(T^{*}M)$
(for  further details of the tensor algebra and properties of the cotangent bundle see for instance \cite{Spi},\cite{1},\cite{2},
also section $2$ in \cite{OP1} introduces
the cotangent bundle and discusses applications of this bundle 
to the description of a relativistic gas.).\\

One important structure - in fact of crucial importance for the development of this work -
is the association of 
 smooth real valued functions on 
$T^{*}M$ induced by 
any smooth contravariant tensor field defined over $(M,g)$. To see this connection, let  
$A$ be 
 any 
smooth $(k,0)$ contravariant tensor field over  $(M,g)$,
then this $A$  induces the smooth 
real valued function 
$\hat A$ via:
\begin{equation}
\hat A :T^{*}M\to R:(x,p)\mapsto \hat A(x,p)=A(x)(p,p,...,p)
\label{AuxF0}
\end{equation}
where $A(x)(p,p,...p)$ 
stands for the value of $A $ at $x$
evaluated on $k$ copies of  $p\in T^{*}_{x}M$. 
As an example, 
let $X$ be  any smooth vector field on $(M,g)$, then 
the real valued function 
$\hat{X}(x,p)$ induced by this $X$  is described by : 

\begin{equation}
\hat{X}(x,p):=X(x)(p)=<p ,X(x)>=X^a(x^\mu)p_a
\label{AuxF1}
\end{equation}
where $<, >$ stands for  the natural pairing between elements of $T^{*}_{x}M$ and $T_{x}M$
and in the last equality we evaluated this pairing relative to  a set of local coordinates
of the background $(M,g)$. 
The maps defined in
(\ref{AuxF0},\ref{AuxF1})
will be frequently employed  in the next sections.
\\

However,  by far the most important structure 
that $T^*M$  acquires from the base manifold, is its natural symplectic structure.
To define this structure. we note
that the  map $\pi$ in (\ref{pro})
induces 
the linear map
$${\pi_{*}}_{(x,p)}: T_{(x,p)}(T^{*}M)\to T_{x}M: L\mapsto {\pi_{*}}_{(x,p)}(L)$$ 
where ${\pi_{*}}_{(x,p)}(L)\in T_{x}M$ is defined so that 
for any smooth $f:M\to R$ we have:

$${\pi_{*}}_{(x,p)}(L)(f)=
\left. L(f \circ  \pi) \right|_{(x,p)}.$$
Since the composition
$f \circ  \pi: T^{*}M\to R$ is smooth,
it follows
that 
$\left. L(f \circ \pi) \right|_{(x,p)}$
is well defined
and thus 
${\pi_{*}}_{(x,p)}(L)$ is also well defined.
We now define the co-vector $\theta$ over $T^{*}M$ via 
 \begin{displaymath}
\theta_{(x,p)}(L) = p_{x}[\pi_{{*}(x,p)}(L)],\qquad L \in T_{(x,p)}T^{*}M.
\end{displaymath}
which is smooth and well defined.
 In  terms of the local coordinates 
$(x^{\mu}, p_{\mu})$,
the form $\theta$ can be written  as:
\begin{displaymath}
\theta_{(x,p)} = \left. p_{\mu}dx^{\mu} \right|_{(x,p)}.
\end{displaymath}
The exterior derivative of $\theta$  defines
the closed two form 
$\Omega=d\theta$
which locally takes the form
\begin{equation}
\Omega_{(x,p)} = \left. dp_{\mu}\right|_{(x,p)} \wedge \left. dx^{\mu} \right|_{(x,p)},
 \label{sfo}
\end{equation}
and this local representation shows
 that $\Omega$ is a non degenerate, closed two-form field on 
$T^{*}M$, i.e.  $\Omega$ serves as a symplectic form over $T^{*}M$.

The bundle $T^{*}M$ equipped with this symplectic form $\Omega$, 
becomes  a smooth symplectic manifold taken as the arena of the Hamiltonian dynamics.
From this perspective, any 
 smooth function $H: T^{*}M\to R$
can serve as a Hamiltonian
and  $\Omega$ determines uniquely
the corresponding Hamiltonian vector field $L_{H}$ on $T^*M$
via 
\begin{equation}
dH=-i_L\Omega=\Omega(,L_{H})
\label{eqHam}
\end{equation}
here $i_L\Omega$ stands for the interior product of $L_{H}$ with $\Omega$.
In the sequel, by the term  
  Hamiltonian flow we mean the flow defined by this Hamiltonian vector field
  $L_{H}$.\\

The symplectic form $\Omega$  also 
defines the Poisson bracket $\{ F, G \}$ for any smooth pair 
$F, G: T^*M\to R$ via
\begin{equation}
\{ F, G \} := dF(L_{G})=\Omega (L_F,L_G).
\label{Poi}
\end{equation}
where $L_{F}, L_{G}$ stand for the Hamiltonian vector fields associated to
$F$, $G$.
As a consequence,
the space of smooth real valued functions,
$C^{\infty}(T^*M, R)$ equipped with the  bracket \{ ,  \} 
becomes a real  Lie algebra.\\

From 
(\ref{Poi}), it follows that a smooth $F:T^*M\to R$ 
is an integral for the flow generated by H, if and only if $F$  and $H$
 are in involution i.e. $\{ F, H \} =0$. More generally, 
 $k$ real valued functions 
 $(F_{1}, F_{2}, ,...,F_{k})$
 are said to  be in involution if 
  $\{ F_{i},F_{j} \} =0$  for all $ i, j =1,2,....,k$.
  Moreover, they are said to be  independent (resp. dependent) at $(x,p) \in T^*M$, 
   if their differentials $(dF_{1}, dF_{2}, . . . , dF_{k})$ at $(x,p)$  are a set
 of linearly independent ( resp. linearly dependent) forms in  $T^{*}_{(x,p)}(T^*M)$.
 This property holds, if and only if the wedge product $(dF_{1}\wedge dF_{2}\wedge dF_{3}.....\wedge dF_{k})
\Big\vert _{(x,p)}\neq0$ 
(respectively $(dF_{1}\wedge dF_{2}\wedge dF_{3}......\wedge dF_{k})\Big\vert _{(x,p)}= 0$).
 Whenever the $k$ integrals in involution  $(F_{1}, F_{2}, ,...,F_{k})$ are independent, then 
 any connected component of
 the set
 $$\Gamma_{(a_{1}, a_{2},....,a_{k})}=\{ (x,p) \in T^*M /F_{1}(x,p)=a_{1}, F_{2}(x,p)=a_{2},....,F_{k}(x,p)=a_{k} \} 
 $$ 
 if non-empty, defines a $k$-dimensional smooth submanifold of  $T^*M$.
 This submanifold remains
 invariant by the flow generated by the corresponding Hamiltonian vector fields $L_{F_{i}}, i=1,2...k$.\\

 We now state the notion of Liouville-Arnold integrability in Hamiltonian dynamics
 (for additional discussion on this integrability consult ref. \cite{1},\cite{2}):
\begin{definition}
\label{Prop:Mf}
The flow of a Hamiltonian H defined on 
$T^*M$ (or more generally any flow generated by a Hamiltonian  $H$ 
over a $2n$ dimensional  symplectic manifold $\hat M$)
 is said to be integrable, or 
completely integrable in the 
Liouville-Arnold sense, 
if there exist $n$ independent integrals 
$(F_{1}=H, F_{2}, ,...,F_{n})$
of  the flow which are in involution.
\end{definition}

Up to this point, the background 
spacetime metric $g$ has not played any role. However when it exists, it defines a natural 
Hamiltonian function $H$ via
\begin{equation}
H: T^*M\mapsto R:(x, p) \rightarrow H(x,p)={1\over 2} {\hat {g}}(x)(p,p)= {1\over 2} g^{\mu\nu}(x)p_\mu p_\nu
\label{eq3}
\end{equation}
where $g^{\mu\nu}(x)$ are  the contravariant components of $g$ 
 relative to the local coordinates $(x^{1},x^{2},....,x^{n})$ of $(M,g)$.
For this 
$H$, it follows from 
(\ref{eqHam})
that the Hamiltonian 
vector field\footnote{The vector field  $L_{H}$ is also  referred
as the Liouville vector field.} $L_{H}$ in the local canonical coordinates 
$(x^{\mu},p_{\mu})$ takes the  form
\begin{equation}
L_{H}=g^{\mu\nu}p_\nu{\partial\over \partial x^\mu}-{1\over 2}{\partial g^{\alpha\beta}\over \partial x^\mu}p_\alpha p_\beta{\partial\over \partial p_\mu}
\label{eq4}
\end{equation}
and it is easily verified that the projections of the integral curves of this $L_{H}$ on 
the base manifold
describe geodesics on the 
spacetime $(M,g)$.
Due to this property, the flow generated by the Liouville vector field  $L_{H}$ (or equivalently by the $H$ in 
(\ref{eq3})) is referred to as a geodesic flow.
 Since $dH(L_{H})=0$ and  the Lie derivative of $H$ along the flow of $L_{H}$ satisfies $\pounds_{L_{H}}H =0$, 
the geodesic flow defined by any
smooth Lorentzian metric can be timelike, null or spacelike, depending on whether the integral curves of the 
 Liouville vector field $L_{H}$ lie on the hypersurfaces defined respectively by
$H(x^\mu,p_\mu)=-m^2<0, H(x^\mu,p_\mu)=0$ or  $H(x^\mu,p_\mu)=m^{2}>0 $.\\
 
The main purpose of the present paper is to discuss properties of the
timelike component of the geodesic  flow 
for the case where $(M,g)$ corresponds 
to a spatially homogenous and spatially isotropic spacetimes, i.e.
$(M,g)$ belongs to the family of Friedmann-Robertson-Walker (FRW) spacetimes.
In the next section, we set up the geodesic flow on  this family of spacetimes.

\section{Constructing the integrals  of motion}
In this section, we introduce  the family of 
of Friedmann-Robertson-Walker (FRW) spacetimes
and for 
 reasons that will become apparent further below, we describe
this family
by employing two coordinates gauges:
 the spherical  $(t,r, \theta, \phi)$ and the Cartesian\footnote{The reason for employing two coordinate gauges is due to the fact that some of the equations 
 in the main text appear to
 become singular when expressed in spherical coordinates but are perfectly regular  when expressed
 in the Cartesian gauge. Furthermore, some computations become much shorter when performed in the Cartesian gauge.}  
 $(t, x, y, z)$
 so that 
$g$ takes the form

\begin{equation}
g = -dt^2 + a^{2}(t)f^2(r)  \left \{ \begin{array}{ll}
dr^2 + r^2(d\vartheta^2 + \sin^{2} \vartheta d\varphi^2) \\
\\dx^2 + dy^2 + dz^2
\end{array}
\right. 
\label{eq1a}
\end{equation}
where 
\begin{equation}
f(r)=\left(1+{kr^2\over 4}\right)^{-1},\quad r^{2}=x^{2}+y^{2}+z^{2}, \quad k\in\{-1,0,1\}.
\label{eq2a}
\end{equation}

Here $a(t)$ is the scale factor while 
$k$ takes the discrete
value of $\{-1,0,1\}$ depending
upon the curvature of the spatial $t=$const space like hypersurfaces.
The choice $(k=-1)$ corresponds to negative curvature,  $(k=0)$ to zero curvature while  $(k=1)$ corresponds to 
 positive curvature.
The range of the coordinate 
$t$ will be taken in the interval $(0, b)$ with $b>0$ (the case where $b \to \infty$ is not excluded).
For the spherical chart, $r$ takes its values in $(0,\infty)$ while 
$(\theta,\phi)$ take their values in the familiar range
while for the Cartesian chart, $(x,y,z)$ take their values over
$(-\infty,\infty).$\\
 
It should be mentioned that even though  these coordinate gauges 
becoming pathological as $r \to 0$ or $r \to \infty$,
these pathologies do not generate serious problems  
as long as we restrict our attention to the spatially flat, i.e. $k=0$
or the case of closed $k=1$ universe.
However for the case of
$k=-1$, the conformal factor of the spatial metric becomes singular at $r^{2}=4$
and this singularity requires special treatment.
Because of these technicalities,
our analysis
covers the $k=0$ and $k=1$ cases (even though the techniques are
extendable to the $k=-1$ case).
\\

We denote by $T^*M$ the eight dimensional cotangent bundle associate with this 
family of spacetimes, and choose as the  Hamiltonian $H$ 
the function defined in (\ref{eq3}).
For the spherical gauge this $H$ takes the form
\begin{equation}
H(x,p)={1\over 2}\left[
-(p_{t})^{2}+\frac {1} {a^{2}f^{2}}[
(p_{r})^{2}+\frac {(p_{\theta})^{2}}{r^{2}}
+\frac {(p_{\phi})^{2}}{r^{2}sin^{2} \theta}]\right]
\label{eqHr}
\end{equation}
while for the Cartesian gauge 
reduces to:
\begin{equation}
H(x,p)={1\over 2}\left[
-(p_{t})^{2}+\frac {1} {a^{2}f^{2}}[
(p_{x})^{2}+(p_{y})^{2}
+(p_{z})^{2}]\right].
\label{eqHc}
\end{equation} 
The local spherical or cartesian coordinates  $(x^{\mu}, p_{\mu})$ over
$T^*M$ are defined having in mind
that for an arbitrary co-vector $p \in T^{*}_{x}(M)$ the following expansions holds:

$$ p=p_{0}dt+p_{r}dr+p_{\theta}d\theta+p_{\phi}d\phi=p_{0}dt+p_{x}dx+p_{y}dy+p_{z}dz$$\\

We now describe a few properties of the flow
defined  by $H$  via  the proposition:

\begin{proposition}
\label{Prop:Mf}
Let $(M,g)$ a  manifold,
$\xi$ a Killing field of $g$ and let on $T^*M$  the Hamiltonian 
 $H(x,p)={1\over 2} {\hat {g}}(x)(p,p)$, then :\\
 
a)  the real valued function 
\begin{equation}
\hat \xi: T^*M\to R: (x,p)\to \hat \xi(x,p)=p(\xi)=\xi^{\mu}p_{\mu}
\label{eqKF}
\end{equation}
 is an integral of motion in the sense
 $\{ H, \hat \xi \} =0$.\\
 
 b) the Hamiltonian vector field $L_{\hat \xi}$ associated with $\hat \xi$ is described by:
 
 \begin{equation}
L_{\hat \xi}=\xi^{\mu}\frac {\partial }{\partial x^{\mu}}-\frac {\partial \xi^{\alpha}}{\partial x^{\mu}}p_{\alpha}\frac {\partial }{\partial p_{\mu}}.
\label{eqHV}
\end{equation} 

c) If $\xi_{i}$ and $\xi_{j}$ are two linearly independent Killing vector fields of $g$ and $\hat \xi_{i}$ 
and $\hat \xi_{j}$
the corresponding functions as in (\ref{eqKF}), then their Poisson bracket satisfies:

 \begin{equation}
\{ \hat \xi_{i},\hat \xi_{j} \}=\Omega (L_{\hat \xi_{i}},L_{\hat \xi_{j}})=[\xi_{i}, \xi_{j}]^{\alpha}{}p_{\alpha}.
\label{PoiCon}
\end{equation}
where  $[\xi_{i}, \xi_{j}]$ is the commutator between $\xi_{i}$ and $\xi_{j}$. 
 
 \end{proposition}

\proof To prove a)  we note from the definition of the Poisson bracket in 
(\ref{Poi}), we obtain
$$
\{ \hat \xi, H \} := d \hat \xi (L_{H})=L_{H}(\hat \xi)=[g^{\mu\nu}p_\nu{\partial\over \partial x^\mu}-{1\over 2}{\partial g^{\alpha\beta}\over \partial x^\mu}p_\alpha p_\beta{\partial\over \partial p_\mu}](\xi^{\gamma}p_{\gamma})=
$$
\begin{equation}
=[-\frac {1}{2}\xi^{\nu}\frac {\partial g_{\alpha\beta}}{\partial x^{\nu}}+g_{\nu\beta}\frac {\partial \xi^{\nu}}{\partial 
x^{\alpha}}]p^{a}p^{\beta}=0
\label{Pro}
\end{equation}
where in the third equality we used the local representation of the Hamiltonian vector
field $L_{H}$ defined in (\ref{eq4}), and the local representation of $\hat \xi$ defined 
in (\ref{eqKF}) and the last equality follows from the fact that $\xi$ satisfies the Killing equation.\\
To prove b), we return to 
(\ref{eqHam}),
 replace $H$ for $\hat \xi$, 
 use local canonical coordinates $(x^{\mu}, p_{\mu})$ to express $L_{\hat \xi}$
in the form
$L_{\hat \xi}=L^{\mu} \frac {\partial }{\partial x^{\mu}}+{\hat L}_{\mu}\frac {\partial }{\partial p_{\mu}}$
and compare:
\begin{equation}
\xi^{\mu}dp_{\mu}+\frac {\partial \xi^{\mu}}{\partial \xi^{\alpha}}p_{\mu}dx^{\alpha}=dx^{\alpha}(L_{\hat \xi})dp_{\alpha}-
dp^{\alpha}(L_{\hat \xi})dx^{\alpha}.
\label{Pro1}
\end{equation}
This  comparison leads immediately to (\ref{eqHV}).\\

To prove c), we appeal to part b) combined with the local representation of $\Omega$
in (\ref{sfo}):
$$
\{ \hat \xi_{i},\hat \xi_{j} \}=\Omega (L_{\hat \xi_{i}},L_{\hat \xi_{j}})=dp_{\mu}(L_{\hat \xi_{i}})dx^{\mu}(L_{\hat \xi_{j}})
-dp_{\mu}(L_{\hat \xi_{j}})dx^{\mu}(L_{\hat \xi_{i}})=[\xi_{i}, \xi_{j}]^{\alpha}{}p_{\alpha}.
$$
\qed

The properties of the geodesic flow described
by  proposition (\ref{Prop:Mf})
will be very useful further ahead, here
 we only mention that this proposition
 is general in the sense that  holds irrespectively whether
the background metric $g$ has Lorentzian, Riemannian or Semi-Riemannian signature, it 
requires however a special Hamiltonian,
namely the one defined  in eq. (\ref{eq3}).\\

It is worth to mention that the Hamiltonian vector field
$L_{\hat \xi}$ in $(\ref{eqHV})$ originates in the Killing field $\xi$ of the background
metric $g$ and thus the one parameter (in general local)
 group of diffeomorphisms that this $\xi$ generates leaves $g$ invariant.
 In the present context, the
 one parameter (in general local)
 group of diffeomorphisms generated  by $L_{\hat \xi}$
leaves $H$ invariant as expressed by the 
vanishing Poisson bracket of $\hat \xi$ with $H$  since:
$$
0=\{ \hat \xi, H \} := dH (L_{\hat \xi})=L_{\hat \xi}(H)= \pounds_{L_{\hat \xi}}H.
$$\\
In this work, the vector field 
$L_{\hat \xi}$ over 
$T^*M$ has been introduced as the unique Hamiltonian vector field 
defined by the real valued function in
(\ref{eqKF}) which in turn is uniquely determined by 
the Killing field $\xi$ admitted by the background metric $g$.
In the  approach of 
 ref. \cite{OP1},
 the field $L_{\hat \xi}$ 
 has been defined via a 
 different route.
They considered a one parameter group of diffeomorphisms (not necessary isometries) acting on $(M,g)$
generated by (a not necessary Killing) vector $\xi$
and subsequently lifted this group
to the bundle
$T^*M$.
Using this lifted group of diffeomorphisms
 they constructed the infinitesimal generator.
 Interestingly, the resulting
 generator coincides with the vector field $L_{\hat \xi}$ introduced in this work\footnote{This coincidence is not an accident.
 Notice that the parts b) and c) of the proposition (\ref{Prop:Mf}) hold irrespectively whether $\xi$
 is Killing or not. Therefore, starting from a smooth vector field $\xi$ on $(M,g)$,
 we first introduce the smooth real valued function $\hat \xi$ defined
 on $T^*M$ and then using the symplectic form we construct the Hamiltonian vector field 
 $L_{\hat \xi}$. This  $L_{\hat \xi}$ generates (at least locally) a one parameter group of diffeomorphisms
 acting upon $T^*M$ which in essence 
 is the same set of operations employed in the ref. \cite{OP1}. }.
 \\

Clearly, the proposition (\ref{Prop:Mf}) implies that
the flow generated by symmetric metrics admit integrals of motion.
For the case of the FRW metrics, 
an integration of the Killing equations $L_\xi g=0$, yields 
 the following set of linearly independent Killing vector fields:
$$
\xi_{(1)}=y \frac{\partial}{\partial z}-z \frac{\partial}{\partial y}=-sin\phi \frac {\partial }{\partial \theta}-cot\theta cos\phi \frac {\partial }{\partial \phi},$$
$$
\xi_{(2)}=z \frac{\partial}{\partial x}-x \frac{\partial}{\partial z}=
cos\phi \frac {\partial }{\partial \theta}-cot\theta cos\phi \frac {\partial }{\partial \phi},
$$
$$
\xi_{(3)}=x \frac{\partial}{\partial y}-y \frac{\partial}{\partial x}=\frac {\partial }{\partial \phi},
$$\\
 $$
\eta_{(1)}=
\left( 1 + \frac{k}{4}(x^2 - y^2 - z^2) \right)\frac{\partial}{\partial x}
 + \frac{k}{2} x y\frac{\partial}{\partial y} + \frac{k}{2} x z\frac{\partial}{\partial z}=
\nonumber
 $$
 $$
 = \left( 1 + \frac{k}{4} r^2 \right)\sin\vartheta\cos\varphi\frac{\partial}{\partial r}
 + \frac{1}{r}\left( 1 - \frac{k}{4} r^2 \right)
 \left( \cos\vartheta\cos\varphi\frac{\partial}{\partial\vartheta}  
  - \frac{\sin\varphi}{\sin\vartheta}\frac{\partial}{\partial\varphi} \right),
$$\\\
$$
\eta_{(2)} = \frac{k}{2} x y\frac{\partial}{\partial x}
 + \left( 1 - \frac{k}{4}(x^2 -y^2 +z^2) \right)\frac{\partial}{\partial y}
 + \frac{k}{2} y z\frac{\partial}{\partial z}=
\nonumber
 $$
 $$
 = \left( 1 + \frac{k}{4} r^2 \right)\sin\theta\sin\phi\frac{\partial}{\partial r}
 + \frac{1}{r}\left( 1 - \frac{k}{4} r^2 \right)
 \left( \cos\theta\sin\phi\frac{\partial}{\partial\theta}  
  + \frac{\cos\phi}{\sin\theta}\frac{\partial}{\partial\phi} \right),
  $$\\\

$$
\eta_{(3)} = \frac{k}{2} x z\frac{\partial}{\partial x} + \frac{k}{2} y z\frac{\partial}{\partial y}
 + \left( 1 - \frac{k}{4}(x^2 +y^2 -z^2 )\right)\frac{\partial}{\partial z}=
\nonumber
 $$
 $$
  = \left( 1 + \frac{k}{4} r^2 \right)\cos\theta\frac{\partial}{\partial r}
 - \frac{1}{r}\left( 1 - \frac{k}{4} r^2 \right)\sin\theta\frac{\partial}{\partial\theta},
$$\\\
 
 where we expressed those fields relative to both spherical coordinates 
  $(r,\theta,\phi)$ and to Cartesian\footnote{We have chosen to express the Killing fields in 
  both spherical and Cartesian components since some of computations
  using Cartesian gauge
  appear much shorter.}
   $\vec{x} = (x,y,z)$ coordinates (of course these sets are 
  related via  $(x,y,z)= r(\sin\theta\cos\phi,\sin\theta\sin\phi,\cos\theta).$ 
 The commutators\footnote{For a compact representation of the commutation relations
(\ref{Eq:ComRelBis}), as well as for 
the representation of the six spatial Killing fields see eqs $(2-6)$ of ref.\cite{O4}.}.
 between these Killing fields read;
\begin{equation}
[\xi_{(1)},\xi_{(2)}] = -\xi_{(3)},\qquad
[\xi_{(1)},\eta_{(2)}] = -\eta_{(3)},\qquad
[\eta_{(1)},\eta_{(2)}] = -k\xi_{(3)},
\label{Eq:ComRelBis}
\end{equation}
and cyclic permutations.
For later use, we record the following brackets:

$$[\xi_{(1)},\eta_{(3)}] =\eta_{(2)},\quad
[\xi_{(2)},\eta_{(1)}] = -\eta_{(3)},\quad
[\xi_{(2)},\eta_{(3)}] = -\eta_{(1)},\quad
$$
\begin{equation}
[\xi_{(3)},\eta_{(1)}] = -\eta_{(2)},\quad
[\xi_{(3)},\eta_{(2)}] =\eta_{(1)}.
\label{CB}
\end{equation}

By appealing to the proposition (\ref{Prop:Mf}) combined 
 with 
 (\ref{AuxF1}), it follows that the six functions:
\begin{equation}
\hat \xi_{(i)}(x,p)
=\xi^{\mu}_{(i)}(x)p_{\mu}, \quad \hat \eta_{(i)}(x,p),
=\eta^{\mu}_{(i)}(x)p_{\mu} \quad i=1,2,3,
\label{ComI}
\end{equation}
are 
 Poisson commuting with the Hamiltonian $H$ in
(\ref{eqHr}) (or the equivalent form of $H$ in (\ref{eqHc})). 
However by appealing to parts b) and c) of proposition (\ref{Prop:Mf})
and the algebra of the commutators
in (\ref{Eq:ComRelBis}), it follows that 
these six integrals fail to commute amongst themselves. In fact we have
the following expressions for the Poisson brackets: 
  \begin{equation}
\{\hat \xi_{(1)}, \hat \xi_{(2)}\} = -\hat \xi_{(3)},\qquad
\{\hat \xi_{(1)},\hat \eta_{(2)}\} = -\hat \eta_{(3)},\qquad
\{\hat \eta_{(1)},\hat \eta_{(2)}\} = -k\hat \xi_{(3)},
\label{Eq:ComRB}
\end{equation} 
 modulo cyclic permutations.
Even though the integrals $\hat \xi_{i}, \hat \eta_{i}, i=(1, 2, 3)$,
in (\ref{ComI})
fail to Poisson commute amongst themselves, one expects that a combination 
of them may  yield integrals with the desired properties. However, it is not clear how to combine these integrals 
in a manner that they yield new Poisson commuting integrals.  In order 
 to resolve this issue, 
in the next section we turn our attention to the analysis of the Hamilton-Jacobi equation.
But before we do so, we consider some particular combinations
of the integrals in 
 (\ref{ComI}) that will be useful in the next section.\\

At first we consider the  azimuthal component $L_{3}$ {and the magnitude
of the angular momentum $L^{2}$ 
defined by
\begin{equation}
L_{3}(x,p)=\hat \xi(x,p)=\xi^{\mu}_{(3)}p_{\mu}=xp_{y}-yp_{x}=p_{\phi}
\label{F1}
\end{equation}

$$
L^2=\hat{\xi}_{(1)}(x^\mu,p_\mu)^2+\hat{\xi}_{(2)}(x^\mu,p_\mu)^2+\hat{\xi}_{(3)}(x^\mu,p_\mu)^2
$$
\begin{equation}
=({\xi}_{(1)}^\mu p_\mu)^2+({\xi}_{(2)}^\mu p_\mu)^2+({\xi}_{(3)}^\mu p_\mu)^2
=p_\theta^2+\frac{p_\phi^2}{\sin^2 \theta}.
\label{F2}
\end{equation}
On the other hand, using the functions
$\hat{\eta}_{(i)}$, we construct:
  $$
K^2=\hat{\eta}_{(1)}(x^\mu,p_\mu)^2+\hat{\eta}_{(2)}(x^\mu,p_\mu)^2+\hat{\eta}_{(3)}(x^\mu,p_\mu)^2
$$
\begin{equation}
=({\eta}_{(1)}^\mu p_\mu)^2+({\eta}_{(2)}^\mu p_\mu)^2+({\eta}_{(3)}^\mu p_\mu)^2
\label{F3}
\end{equation}
and a straightforward but long algebra, using the Cartesian representation of the generators
yields
$$
K^2=p_{x}^{2} +p_{y}^{2}+p_{z}^{2} + \frac {k^{2}}{16} [p_{x}^{2}+p_{y}^{2}+p_{z}^{2}][x^{2}+y^{2}+z^{2}]^{2}+
$$
$$
+\frac {k}{2}[p_{x}^{2}(x^{2}-y^{2}-z^{2})+p_{y}^{2}(y^{2}-z^{2}-x^{2})+p_{z}^{2}(z^{2}-x^{2}-y^{2})]+
$$
$$
+2k[xyp_{x}p_{y}+xzp_{x}p_{z} +yzp_{y}p_{z}].
$$
The  representation of the right hand side in terms of the spherical coordinates
is long and not very revealing.
Interestingly however,
the combination 
$K^2+kL^2$ has the following  simple form:

\begin{equation}
K^2+kL^2=        
\frac{1}{f(r)^2}\left[p_x^2+p_y^2+p_z^2]\right]
=\frac{1}{f(r)^2}\left[p_r^2+\frac{p_\theta^2}{r^2}+\frac{p_\phi^2}{r^2\sin^2\theta}\right].
\label{SF3}
\end{equation}
 
As we shall see in the next section, the right hand sides of
(\ref{F1}, \ref{F2}, \ref{SF3}) appear naturally in the Hamilton-Jacobi equation.
This remarkable property is consequence of the complete separability of the Hamilton-Jacobi equation on an
FRW background expressed in the spherical gauge and this problem is analyzed in the next section.

\section{Separability of the Hamilton-Jacobi equation}

The Hamilton-Jacobi equation for the Hamiltonian described by
(\ref{eq3}) has the form\footnote{Note that the form of the Hamilton-Jacobi equation in (\ref{HJ})
assumes
that $H$ in 
(\ref{eq3}), is normalized according to $H(x,p)=-\frac {m^{2}}{2}$.}

 \begin{equation}\label{eq6}
g^{\mu\nu}(x){\partial S\over \partial x^\mu}{\partial S\over\partial x^\nu}=-m^2,\quad p_{\mu}=\frac {\partial S}{\partial x^{\mu}}
\label{HJ}
\end{equation}
and for the case of the spherical gauge defined  in (\ref{eq1a}) reduces to:
$$
-a^2(t)\left({\partial S\over\partial t}\right)^2
+m^2a^2(t)+
$$
\begin{equation}
+{1\over f^2(r)}\left[\left({\partial S\over\partial r}\right)^2+
{1\over r^2}\left({\partial S\over\partial \theta}\right)^2+
{1\over r^2\sin^2\theta}\left({\partial S\over\partial \phi}\right)^2\right]=0.
\label{eq7}
\end{equation}\\

 The  ansatz 
\begin{equation}
S(t,r,\theta,\phi)=S_t(t)+S_r(r)+S_\theta(\theta)+S_\phi(\phi)
\label{eq8}
\end{equation}
implies
\begin{equation}\label{eq9}
\left({\partial S_{t}(t)\over\partial t}\right)^2=m^2+{\lambda^2\over a^2(t)},
\end{equation}
\begin{equation}\label{eq10}
\left({\partial S_{r}(r)\over\partial r}\right)^2=\lambda^2{f^2(r)}-{a^{2}_\theta\over r^2},
\end{equation}
\begin{equation}\label{eq11}
\left({\partial S_{\theta}(\theta)\over\partial \theta}\right)^2=a^{2}_\theta-{a_\phi^2\over \sin^2\theta},
\end{equation}
\begin{equation}\label{eq12}
\left({\partial S_{\phi}(\phi)\over\partial \phi}\right)^2=a^2_\phi,
\end{equation}
where  $(\lambda^2,a^{2}_\theta,a_\phi^2)$ are separations constants.
 The complete separability\footnote{A referee kindly pointe out to us that
the separability of the Hamilton-Jacobi equation (\ref{HJ})  on an FRW background
has been suggester in ref. \cite{PAD}. However beyond this suggestion no further
analysis has been pursued in \cite{PAD} on the implications of this separability.}
 the 
of (\ref{eq7}) is welcomed and implies
that 
 the function 
$$
S(t,r,\theta,\phi,m,\lambda,a_\theta,a_\phi)\!=\!\!\int^{t}\!\!\!\sqrt{m^2+{\lambda^2\over a^2(t')}}dt'+
\int^{r}\!\!\! \sqrt{\lambda^2 f^2(r')-{a^{2}_\theta\over {r'}^2}}dr'\qquad\qquad
$$
\begin{equation}\label{eq13}
\quad+\int^{\theta}\!\!\!\sqrt{a^{2}_\theta-{a_\phi^2\over \sin^2\theta'}}d\theta'
+a_\phi \phi
\end{equation}
generates canonical transformations that
trivialize the Hamiltonian (see for example  \cite{1}, \cite{3}). Indeed,
defining $(Q^{\mu}, P_{\mu})$  on $T^* M$ via
$$
P_0=m,\quad P_1=\lambda^2,\qquad P_2=a_\theta,\quad P_3=a_\phi
$$
$$
Q^1={\partial S\over\partial m},\qquad
Q^2={\partial S\over\partial \lambda^2},\qquad
Q^3={\partial S\over\partial a_\theta},\qquad
Q^4={\partial S\over\partial a_\phi}
$$
 it is easy to see that the set $(Q^{\mu}, P_{\mu})$ defines new canonical coordinates
 and relative to these new coordinates the Hamiltonian is 
  trivial. Moreover, since by definition $p_{\mu}=\frac {\partial S}{\partial x^{\mu}} \Longrightarrow p^{\mu}=m\frac {dx^{\mu}}{d\tau}=g^{\mu\nu}p_{\nu},$ we find in an almost labor free manner
 that the first integrals describing
   causal geodesics on an FRW background
 take the form:
 $$
(\frac {dt}{d\tau})^{2}=m^2+{\lambda^2\over a^2(t)},\quad (\frac {dr}{d\tau})^{2}=
\frac {1}{a^{4}f^{2}}[\lambda^2-\frac {{a^{2}_\theta}}{ f^{2}r^2}],
$$
\begin{equation}
(\frac {d\theta}{d\tau})^{2}=\frac {1}{a^{2}f^{2} r^{2}}[a^{2}_\theta-{a_\phi^2\over \sin^2\theta}],\quad
(\frac {d\phi}{d\tau})^{2}=\frac {a^{2}_{\phi}}{a^{2}f^{2}r^{2}\sin^{2}\theta}
\label{FIRI}
\end{equation}
where 
we absorbed the mass parameter $m$  in the redefinition of the proper time $\tau$. \\

In this section we do not analyze 
the above integrals of geodesic motion, nor
we explore
the implications
of the new canonical chart $(Q^{\mu}, P_{\mu})$ over 
 $T^*M$ although both of these issues are 
 worth of further analysis.
 The purpose of this section is 
to settle the issue regarding the number of integrals in involution with 
themselves and the Hamiltonian $H$.\\
In order to see how the separability of 
(\ref{eq7}) settles this problem, we return 
to (\ref{eq9}-\ref{eq12}) and first we solve the separation constants
 $(\lambda^2,a^{2}_\theta,a_\phi^2)$ 
in terms of the canonical momenta
$p_{\mu}=\frac {\partial S}{\partial x^{\mu}}$. After some manipulations, we 
get:
\begin{equation}
-m^{2}=-(p_{t})^{2}+\frac {\lambda^{2}}{a^{2}}=-(p_{t})^{2}+\frac {1}{a^{2}f^{2}}[p_{r}^{2}+\frac {p_{\theta}^{2}}{r^{2}}+
\frac {p_{\phi}^{2}}{r^{2}sin^{2}\theta}]
\label{eq9A}
\end{equation}
\begin{equation}
\lambda^{2}=
\frac {1}{f^{2}}[p_{r}^{2}+\frac {p_{\theta}^{2}}{r^{2}}+
\frac {p_{\phi}^{2}}{r^{2}sin^{2}\theta}]
\label{eq10A}
\end{equation}
\begin{equation}\label{eq11A}
a_{\theta}^{2}=p_{\theta}^{2}+\frac {p_{\phi}^{2}}{sin^{2}\theta},\quad a^{2}_{\phi}=p^{2}_{\phi}
\label{eq11A}
\end{equation}
However, the right hand sides of
(\ref{eq9A}-\ref{eq11A})
in combination
to
 (\ref{F1}, \ref{F2}, \ref{SF3}),
suggest to  
introduce 
the  functions:
 \begin{equation}
F_{i}: T^*M\to R: (x,p)\to F_{i}(x,p), \quad i=1,2,3,4
\label{IntM}
\end{equation}
with:
$$
F_{1}(x,p)=2H(x,p),\quad F_{2}(x,p)=L^{2}(x,p),\quad F_{3}(x,p)=L_{3}(x,p),
$$
\begin{equation}
F_{4}(x,p)
=K^{2}(x,p) +kL^{2}(x,p).
\label{IntMS}
\end{equation}
Expressing these functions in terms of the local coordinates $(x^{\mu},p_{\mu})$,
it is seen that
relations (\ref{eq9A}-\ref{eq11A})
just describe 
the level surfaces of $F_{i}$, i.e.:
$$
F_{1}(x,p)=2H(x,p)=-m^{2},\quad F_{4}(x,p)
=K^2(x,p)+kL^2(x,p)=\lambda^{2}
$$
\begin{equation} F_{2}(x,p)=L^{2}(x,p)=a^{2}_{\theta},\quad 
F_{3}(x,p)=L_{\phi}(x,p)=a_{\phi}.
\label{eqCon1}
\end{equation} \\
We now show that the functions $(F_{1}=2H, F_{2}, F_{3}, F_{4})$ are a set of
Poisson commuting integrals.
For that,
it is convenient to construct first the associated Hamiltonian vector fields
denoted  by $L_{F_{i}}$.
 Clearly
$L_{F_{1}}=L_{H}$ is just the
Louville vector field described in  (\ref{eq4}) while $L_{F_{3}}=L_{L_{3}}=\frac {\partial}{\partial {\phi}}$.
On the other hand, a computation 
based on 
(\ref{eqHam}) shows that 
the Hamiltonian field associated with $L^{2}$ has the form:
$$
L_{F_{2}}:=L_{L^{2}}=\xi_{(1)}^{k} p_{k}L_{\hat \xi_{(1)}}+
\xi_{(2)}^{k}p_{k}L_{\hat \xi_{(2)}}+
\xi_{(3)}^{k}p_{k}L_{\hat \xi_{(3)}}=
$$
$$
=a_{1}L_{\hat \xi_{(1)}}+
a_{2}L_{\hat \xi_{(2)}}
+a_{3}L_{\hat \xi_{(3)}},\quad a_{i}=\xi_{(i)}^{\mu}p_{\mu},\quad i=1,2,3,
$$
where the second representation of $L_{L^{2}}$ will be used shortly.
A similar computation shows that the Hamiltonian field $L_{K^{2}}$
has the form:
$$
L_{F_{2}}:=L_{K^{2}}=\eta_{(1)}^{k} p_{k}L_{\hat \eta_{(1)}}+
\eta_{(2)}^{k}p_{k}L_{\hat \eta_{(2)}}+
\eta_{(3)}^{k}p_{k}L_{\hat \eta_{(3)}}=$$
$$
=\hat a_{1}L_{\hat \eta_{(1)}}+
\hat a_{2}L_{\hat \eta_{(2)}}
+\hat a_{3}L_{\hat \eta_{(3)}}, \quad
\hat a_{i}=\eta_{(i)}^{\mu}p_{\mu},\quad i=1,2,3
$$
Using these Hamiltonian vector fields,
we get as an immediate consequence
of the part $a)$ of the proposition
 (\ref{Prop:Mf})  that
$\{ H, F_{i} \}=0$
for all $i \in {1,2,3}$.
On the other hand, by employing
the local canonical coordinates $(x^{\mu},p_{\mu})$,
a short computation shows that: $\{ L_{3}, L^{2} \}=
\{ L_{3}, K^{2} \}=0$. We are therefore left
to investigate whether the Poisson bracket between $L^{2}$ and $K^{2}$
is also vanishing.
By appealing to the definition of the 
Poisson bracket in 
(\ref{Poi}) it follows that
$$ 
\{ L^{2}, K^{2} \}= \Omega (L_{L^{2}},L_{K^{2}})=a_{1} \hat a_{2}[\xi_{(1)}, \eta_{(2)}]^{a}p_{a}+
a_{1}\hat a_{3}[\xi_{(1)}, \eta_{(3)}]^{a}p_{a}+$$
$$a_{2}\hat a_{1}[\xi_{(2)}, \eta_{(1)}]^{a}p_{a}+a_{2}\hat a_{3}[\xi_{(2)}, \eta_{(3)}]^{a}p_{a}+
a_{3}\hat a_{1}[\xi_{(3)}, \eta_{(1)}]^{a}p_{a}+$$
$$a_{3}\hat a_{2}[\xi_{(3)}, \eta_{(2)}]^{a}p_{a}=0$$
where the last equality comes from taking into account the commutators 
between $\xi$ and $\eta$ listed earlier on.

Thus, the functions\footnote{It is important 
to mention here that the functions
$(F_{1}=2H, F_{2}, F_{3}, F_{4})$ 
defined
in $(\ref{IntM}-\ref{IntMS})$
are defined over the entire bundle
$T^{*}M$. Locally defined integrals cannot be used
to conclude 
Liouville-Arnold integrability.
They are required to be extended as smooth integrals over the entire bundle.}
$(F_{1}=2H, F_{2}, F_{3}, F_{4})$ 
in $(\ref{IntM}-\ref{IntMS})$
are Poisson commuting amongst themselves.
Using these integrals, we now introduce the sets\footnote{Here for convenience, we have chosen
the constants determining the level surfaces of the $F_{i}$ to be the same as those appearing in the 
separated Hamilton-Jacobi equation
(\ref{eq9A}-\ref{eq11A}).}:
$$
\Gamma_{(m,\lambda^{2},a_{\theta}^{2},a_{\phi})}=\{(x,p)\in T^{*}_{x}M, \ F_{1}(x,p)=2H(x,p)=-m^{2},
$$
\begin{equation}
F_{2}(x,p)=a_{\theta}^{2},\quad F_{3}(x,p)=a_{\phi},\quad F_{4}(x,p)=\lambda^{2} \}. 
\label{INV}
\end{equation}
i.e. these sets are the common level surface of the integrals.

In the next section, we investigate 
whether
these sets are  non empty
and if non empty whether
they
define a four dimensional smooth 
submanifolds\footnote{As it turns out, some of these sets are singular submanifolds
in the following sense:
 they are 
regions of 
$T^{*}M$ where either one or more of the $dF_{i}$ vanishes,
or two or more of the $dF_{i}$ become linearly dependent.}
of $T^{*}M$.\\

\section{The structure of the Invariant submanifolds}

As we have mentioned 
in section $1$, the integrals $(F_{1}=2H, F_{2}, F_{3}, F_{4})$
are considered to be independent
at $(x,p)$ provided
$(dF_{1}\wedge dF_{2}\wedge dF_{3}\wedge dF_{4})\Big\vert _{(x,p)} \neq 0$.
Since $F_{2}=L^{2}$ and $F_{4}=K^{2}+kL^{2}$, it
 is sufficient to examine the product
$(dF_{1}\wedge dF_{2}\wedge dF_{3}\wedge dK^{2})$
instead of $(dF_{1}\wedge dF_{2}\wedge dF_{3}\wedge dF_{4})\Big\vert _{(x,p)} \neq 0$.
From the definition,
 we find that at any 
 $(x,p)$
: 
  \begin{equation}
dF_{1}=2dH=2[g^{\mu\nu}p_{\nu}dp_{\mu} +\frac {1}{2}\frac {\partial g^{\alpha\beta}}{\partial x^{\mu}}p_{\alpha}p_{\beta}dx^{\mu}]
\label{dif1}
\end{equation}
 $$
 dF_{2}=dL^{2}=2(\xi_{(1)}^{\nu}p_{\nu})[\xi_{(1)}^{\mu}dp_{\mu}+\frac {\partial \xi_{(1)}^{\mu}}{\partial x^{\alpha}}p_{\mu}dx^{\alpha}] $$
$$+2(\xi_{(2)}^{\nu}p_{\nu})[\xi_{(2)}^{\mu}dp_{\mu}+\frac {\partial \xi_{(2)}^{\mu}}{\partial x^{\alpha}}p_{\mu}dx^{\alpha}]
$$
\begin{equation}
+2(\xi_{(3)}^{\nu}p_{\nu})[\xi_{(3)}^{\mu}dp_{\mu}+\frac {\partial \xi_{(3)}^{\mu}}{\partial x^{\alpha}}p_{\mu}dx^{\alpha}]
\label{dif2}
\end{equation}
\begin{equation}
dF_{3}=dL_{3}=
[\xi_{(3)}^{\mu}dp_{\mu}+\frac {\partial \xi_{(3)}^{\mu}}{\partial x^{\alpha}}p_{\mu}dx^{\alpha}]
\label{dif3}
\end{equation}
$$
dK^{2}=2(\eta_{(1)}^{\nu}p_{\nu})[\eta_{(1)}^{\mu}dp_{\mu}+\frac {\partial \eta_{(1)}^{\mu}}{\partial x^{\alpha}}p_{\mu}dx^{\alpha}] $$
$$+2(\eta_{(2)}^{\nu}p_{\nu})[\eta_{(2)}^{\mu}dp_{\mu}+\frac {\partial \eta_{(2)}^{\mu}}{\partial x^{\alpha}}p_{\mu}dx^{\alpha}]
$$
\begin{equation}
+2(\eta_{(3)}^{\nu}p_{\nu})[\eta_{(3)}^{\mu}dp_{\mu}+\frac {\partial \eta_{(3)}^{\mu}}{\partial x^{\alpha}}p_{\mu}dx^{\alpha}].
\label{dif4}
\end{equation}\\
Using these formulas, a straightforward evaluation of 
$(dF_{1}\wedge dF_{2}\wedge dF_{3}\wedge dK^{2})$
yields a long expression which is not very illuminating
in identifying regions where 
$(dF_{1}\wedge dF_{2}\wedge dF_{3}\wedge dK^{2})$
is non vanishing.
However, 
based on the structure of 
($\ref{dif1}-\ref{dif4}$),
a few comments are helpful
in addressing that problem.\\

Firstly we note that $(\ref{dif1})$ implies
that 
$dH(x,p)\neq0$
at any $(x,p)\in T^{*}(M)$, since 
$dx^{\mu}$ and $dp_{\mu}$ are linearly independent
and the possibility that all $p_{\mu}$ vanish contradicts the normalization condition\footnote{
As we have mentioned  in the introduction,
this work focus on the timelike component of the geodesic flow.
The case of the null component although physically is very important,
 needs further analysis. For instance  for the null case $dH(x,p)$ admits zeros  
and these zeros require special treatment.}
$H(x,p)=-m^{2}$.
As far as 
the forms
$(dF_{2}, dF_{3}, dK^{2})$
are concerned, 
 we observe that they
are determined by the Killing fields
$\xi_{(i)}$ and $\eta_{(i)}$
and here we recall
that any  non trivial\footnote{A non trivial Killing field $\xi$ is a field which is not identically 
vanishing over an open region of interest.} Killing  field $\xi$ cannot 
simultaneously
satisfy\footnote{Here 
$d\xi(q)=0$ is a short hand notation that all partial derivatives $\frac {\partial \xi ^{\mu}}{\partial x^{\nu}}$ 
are vanishing at $q$.}
$\xi(q)=d\xi(q)=0$
at any $q \in (M,g)$
 (for a proof of this property, see for instance \cite{Wald}).
This property  implies that
the terms within the brackets  in 
the right hand sides of (\ref{dif2}-\ref{dif4}) cannot become
individually zero
(recall not all $p_{\mu}$ can be zero)
which means that 
$dF_{3}$ is non zero over 
$T^{*}(M)$.
However, the factors 
$\hat \xi_{(i)}(x,p)=\xi_{(i)}^{\mu}(x)p_{\mu}$ and  
 $\hat \eta_{(i)}(x,p)=\eta_{(i)}^{\mu}(x)p_{\mu}$
in (\ref{dif2},\ref{dif4}),
imply that
$(dF_{2}, dK^{2})$ can vanish.
For instance, 
$dF_{2}$ 
vanishes
over regions  where 
all $\hat \xi_{(i)}(x,p)=\xi_{(i)}^{\mu}(x)p_{\mu}$ vanish,
while
$dK^{2}$ vanishes
when all
 $\hat \eta_{(i)}(x,p)=\eta_{(i)}^{\mu}(x)p_{\mu}$ become zero.
 Moreover, 
 $dF_{2}=dL^{2}$ and $dK^{2}$ 
 vanish simultaneously 
 at those $(x,p)$ satisfying
 $\hat \xi_{(i)}(x,p)=
 \hat \eta_{(i)}(x,p)=0$.
Further ahead we show that 
all these possibilities
exist.\\

 Finally, 
 $(dF_{1}\wedge dF_{2}\wedge dF_{3}\wedge dK^{2})$
can vanish whenever
on 
the level surfaces of the integrals 
exist $(x,p)$  
where at least two of the 
forms $(dH, dF_{2}, dF_{3}, dK^{2})$ become linearly dependent.
Clearly, at such points 
  $(dF_{1}\wedge dF_{2}\wedge dF_{3}\wedge dK^{2})$
  vanishes and below we show  this possibility indeed occurs.\\


  With these general remarks in mind, we now study
 the sets
 $\Gamma_{(m,\lambda^{2},a_{\theta}^{2},a_{\phi})}$
 defined
in (\ref{INV}), and at first we consider the 
case where
$m^{2}>0$, 
  $(\lambda^{2}, ~a^{2}_{\theta}) \in (0,\infty)$
and $a_{\phi} \in R$.
Our first task is to investigate whether 
$\Gamma_{(m^{2}>0,~\lambda^{2}>0,~a_{\theta}^{2}>0,~a_{\phi})}$
is non empty. One way to check this property, 
is to solve algebraically
the system:
$F_{1}(x_{0},p_{0})=2H(x_{0},p_{0})=-m^{2}>0, 
F_{2}(x_{0},p_{0})=a^{2}_{\theta}>0, F_{3}(x_{0},p_{0})=a_{\phi}, F_{4}(x_{0},p_{0})=\lambda^{2}>0$
for an initial point  $(x_{0}, p_{0})$ on the bundle.
If this system admits  an $(x_{0}, p_{0})$ as a solution,
then we  propagate 
$F_{1}(x_{0},p_{0})=2H(x_{0},p_{0}), 
F_{2}(x_{0},p_{0}), F_{3}(x_{0},p_{0}), F_{4}(x_{0},p_{0})$
along the integral curves of the Hamiltonian vector fields
$L_{F_{i}}$. 
The maximal connected components consisting of all $(x,p)$ lying on these integral curves 
 define the set
$\Gamma_{(m^{2}>0,~\lambda^{2}>0,~a_{\theta}^{2}>0,~a_{\phi})}$
which by construction is non empty.\\
Although this seems to be a reasonable way to proceed, from the 
 practical point of view, it is a difficult algorithm to implement
 since in general the integrals $F_{i}$ are complicate expressions.
An alternative way to proceed is to 
use the local representation
of the integrals and study 
the individual levels  
on the planes $(t,p_{t})$, $(r,p_{r})$, etc,
along the same lines as 
pursued in
\cite{OP1}.\\
Via this approach, 
at first, we consider the levels of 
the integral $F_{1}(x,p) = 2H(x,p) $
which  
satisfy 
$F_{1}(x,p) = 2H(x,p) = -m^{2}<0$.
In view of 
(\ref{eq9A})
these levels consists of the points $(t,p_{t})$ 
satisfying
\begin{equation}
p^{2}_{t} -\frac {\lambda^{2}}{a^{2}(t)}=m^{2}.
\label{Re1}
\end{equation}
The graph of this equation
 on the $(t,p_{t})$-plane   
depends upon the scale factor
$a(t)$ and the parameter $\lambda^{2}$.
If $\lambda^{2}>0$, then 
the levels consists of two disconnected 
regular branches 
described by 
\begin{equation}
T_{\pm}= \{(t,p_{t}) \in T^{*}M, \ t\in (0,b), b>0,  p_{t}=\pm [m^{2}+\frac {\lambda^{2}}{a^{2}(t)}]^{\frac {1}{2}} \}.
\label{InvT}
\end{equation}
 For an open universe, assuming a power law behavior,  i.e. 
$a(t)=t^{\gamma}, t\in (0,\infty), \gamma>0$,
asymptotically i.e. as $t \to \infty$, the $p_{t}$ component approaches the rest mass
of the 
particle while at the other extreme, i.e. as $t\to 0$, 
the rest mass becomes irrelevant since $p_{t}$ is dominated by
the particles kinetic energy.
For the case where $a(t)$ describes a closed, recollapsing at a finite time $t_{f}$ universe,
still
the level surfaces consist of a family of open regular curves 
but in this case $p_{t}$ diverges as $t\to0$ and also $t\to t_{f}$.
Notice that in the particular case where $\lambda^{2}=0$,
the levels consist again of two disconnected
characterized by $p_{t}=\pm m$ as can be easily seen from (\ref{InvT}).
In all cases,
as long as $m^{2}\in (0, \infty)$,
the levels of 
$F_{1}(x,p) = 2H(x,p) = -m^{2}<0$ 
consist of open submanifolds diffeomorphic  to 
the real line.
\\

The levels defined by $F_{2}(x,p)$
satisfy 
$F_{2}(x,p)=a^{2}_{\theta}>0$ and 
in the spherical gauge they are consist of 
$(p_{\theta}, \theta)$ obeying:

\begin{equation}
p_{\theta}^{2} +\frac {a^{2}_{\phi}}{sin^{2} \theta}=a^{2}_{\theta},\quad a_{\phi} \in R,
\label{Cir1}
\end{equation}
which is 
an equation 
similar to
(\ref{Re1}) except for the crucial  sign difference
in the second term. 
Equation (\ref{Cir1})
 is the trademark of spherical symmetry and thus it has been encountered 
in other contexts, for example in the analysis of 
the Kepler problem (see for instance \cite{3}), or in the analysis of spherical relativistic systems (see for instance
\cite{OP1}). From these studies, it follows that
as long as  $0<a^{2}_{\phi}<a^{2}_{\theta}$, then
(\ref{Cir1})
 describes  a family of closed 
curves on the $(\theta, p_{\theta})$ plane
that shrinks to the point  $(\theta= \frac {\pi}{2}, p_{\theta}=0)$
occurring when
$a^{2}_{\phi}=a^{2}_{\theta}$.
Therefore,
provided  $0<a^{2}_{\phi} < a^{2}_{\theta}$, the levels  of
the integral $F_{2}(x,p)$ are topologically circles
(and here we left out
particular values of 
$a^{2}_{\phi}$ and  $a^{2}_{\theta}$
that will be analyzed further bellow.)\\

The levels of $F_{3}(x,p)=a_{\phi}$ when expressed in the local 
spherical coordinates take the form $p_{\phi}=a_{\phi}$
and as  long as $a_{\phi}$ is fixed, 
 they  consist  of $(\phi, p_{\phi})$ with
$\phi \in [0, 2\pi]$ and thus define a circle $S^{1}$.\\

Finally, we analyze the levels 
of  the integral 
$F_{4}(x,p)$
that satisfy
$F_{4}(x,p):=K^{2}(x,p)+kL^{2}(x,p)=\lambda^{2}$.
In view of  (\ref{F2}) and (\ref{SF3})
 these levels consist of those 
$(r, p_{r})$ obeying
\begin{equation}
p_{r}^2=\lambda^2{f^2(r)}-{a^{2}_\theta\over r^2}.
\label{INVR}
\end{equation}
For
 $\lambda^{2}>0$ and $a^{2}_{\theta}>0$, 
positivity of the right hand side
demands:
  \begin{equation}
\frac {\lambda^{2}}{a^{2}_\theta} \geq V(r):=\frac {1}{f^{2}r^{2}}=(\frac {1}{r}+\frac{kr}{4})^{2}.
\label{RAR}
\end{equation}
and this inequality 
is satisfied provided
\begin{equation}
r_{-}\leq r \leq r_{+}, \quad r_{\pm}=\frac {2}{k}[{\hat a\pm(\hat a^{2}-k)^{\frac {1}{2}}}], \quad \hat a=\frac {\lambda}{a_{\theta}}
\label{RARC}
\end{equation}
where $r_{\pm}$ are the two roots of the equation
$\hat a=f^{-1}r^{-1}$ which are real and 
distinct provided $\hat  a^{2}:=(\frac {\lambda}{a_{\theta}})^{2}>k$.
For a spatially flat universe, i.e. $k=0$, 
 this inequality 
always hold as a consequence of 
$ \lambda^{2}>0$ and $a^{2}_{\theta}>0$
while 
for a universe with compact spatial sections, i.e. $k=1$,
requires\footnote{We are avoiding to
 address the case of a $k=-1$ since for this case
 the potential
 $V(r)=(rf)^{-2}$ vanishes at $r^{2}=4$, i.e. at the point where the conformal factor
 in (\ref{eq1a}) becomes singular. The case of a spatially hyperbolic universe will be discussed elsewhere.} 
 $\lambda^{2}>a^{2}_{\theta}$.\\
 
 For $k=1$
 and for
 $\lambda^{2}>a^{2}_{\theta}$ 
 so that
$\hat a=f^{-1}r^{-1}$ 
 admits two real distinct roots $r_{\pm}$,
 we find that 
 $r(\tau)$ obeys
\begin{equation}
(\frac {dr}{d\tau})^{2}=
\frac {1}{a^{4}f^{2}}[\lambda^2-\frac {{a^{2}_\theta}}{ f^{2}r^2}]=\frac {a^{2}_{\theta}}{a^{4}f^{2}}[\frac {\lambda^2}{a^{2}_{\theta}}-\frac {1}{ f^{2}r^2}],
\label{DE}
\end{equation}
which shows that
the radial motion is restricted on the closed interval 
$[r_{-}, r_{+}]$ having turning points at $r_{\pm}$.
For the particular values $\lambda^{2}=a_{\theta}^{2}$ the motion becomes circular 
at  $r_{min}=2$ which corresponds to the minima of the potential
$V(r)=(rf)^{-2}$.\\

For  $k=0$,
the potential $V(r)$
in (\ref{RAR}) reduces to  $V(r)=r^{-2}$
which implies that for a given $\hat a=\lambda (a_{\theta})^{-1}$
the motion takes place in $(r_{min}, \infty)$ where $r_{min}$ is the root of 
$r^{2}=a^{2}_{\theta}\lambda^{-2}$.\\

So far, from this analysis
we conclude that the sets:
$\Gamma_{(m^{2}>0,~\lambda^{2}>0,~a_{\theta}^{2}>0,~a_{\phi})}$
are
not empty
and now examine whether
these sets
define smooth four dimensional submanifolds
of
$T^{*}M$.
 To check 
whether this is the case, at first
we evaluate the forms $dF_{i}$ on $\Gamma_{(m^{2}>0,~\lambda^{2}>0,~a_{\theta}^{2}>0,~a_{\phi})}$
and determine whether exist values of the parameters where the forms
$dF_{i}$
are linearly independent. Moreover, we also investigate whether $dF_{2}$ 
or $dF_{4}$ vanish
on $\Gamma_{(m^{2}>0,~\lambda^{2}>0,~a_{\theta}^{2}>0,~a_{\phi})}$.\\

Starting from $dH(x,p)$ in 
(\ref{dif1}) we evaluate 
$dH(x,p)$ on the level sets of the integrals.
In view of the special form 
of $K^{2}+kL^{2}$
shown in
(\ref{SF3}), we find
that
$dH(x,p)$ takes the form:
\begin{equation}
dH(x,p)=\frac {1}{2}[-2p_{t}dp_{t}-\frac {2}{a^{3}}\frac {da}{dt}\lambda^{2}dt+\frac {1}{a^{2}}(dK^{2}+kdL^{2})]
\label{DH}
\end{equation}
which shows that $dH$ is always linearly independent from $dK^{2}$ and $dF_{2}=dL^{2}$
even for the values $\lambda^{2}=0$ or  when the forms 
 $dK^{2}$ and $dL^{2}$ are vanishing at some $(x,p)$.
 This conclusion holds 
  provided that $m^{2}>0$.\\
  
 On the other hand, 
   (\ref{dif2})
  shows that 
  $dF_{2}=dL^{2}$ 
    vanishes at any $(x,p)$ satisfying 
 $\hat \xi_{(i)}(x, p) =0$ for all $i=(1,2,3)$.
 This possibility 
 occurs for the  case
where  $dF_{2}=dL^{2}$ is evaluated on the level surfaces defined by:
  $F_{2}=L^{2}=0$ and below  we examine in detail this case. 
  Here we are interested to
 see whether $dF_{2}=dL^{2}$ is linearly independent
 from  the rest of the forms when it is evaluated on 
$\Gamma_{(m^{2}>0,~\lambda^{2}>0,~a_{\theta}^{2}>0,~a_{\phi})}$.  
Using the representation of $L^{2}$ in the 
 spherical gauge, we obtain:
 \begin{equation}
dF_{2}(x,p)=dL^{2}(x,p)=2p_{\theta}dp_{\theta}-\frac {2a_{\phi}^{2}cos\theta}{sin^{3}\theta}d\theta+2\frac {a_{\phi}}{sin^{2}\theta}dF_{3},
\quad a_{\phi}\in R.
\label{DL}
\end{equation} 
which shows 
that 
$dF_{2}$ are $dF_{3}$ are linearly independent
except for 
 the values  $\theta=\frac {\pi}{2}, p_{\theta}=0$ lying on the
levels of 
$L^{2}$ and $L_{3}$ obeying: $a^{2}_{\theta}=a^{2}_{\phi}$.
For this case
$dF_{2}=dL^{2}$ and $dF_{3}$ fail to be linearly independent
and this occurs when the motion is confined on the equatorial plane.

Finally, we consider
 the form $dF_{4}$ and  since 
$F_{4}(x,p)=K^{2}(x,p)+kL^{2}(x,p)$
it follows from
 (\ref{dif4}) and (\ref{dif2})
that $dF_{4}$
 vanishes at any $(x,p)$ obeying
 $\hat \xi_{(i)}(x,p)=
 \hat \eta_{(i)}(x,p)=0, i\in (1,2,3)$.
 This case will be analyzed further ahead.
 However, evaluating
 $dF_{4}$ on 
 $\Gamma_{(m^{2}>0,~\lambda^{2}>0,~a_{\theta}^{2}>0,~a_{\phi})}$  
 in view of 
 (\ref{SF3}), we find
\begin{equation}
dK^{2}=\frac {2p_{r}}{f^{2}}dp_{r}+[p^{2}_{r}\frac {kr}{f}-\frac{2}{r^{3}f}(1-\frac {kr^{2}}{4})a^{2}_{\theta}]dr+
(\frac {1}{f^{2}r^{2}}-k)dL^{2}
\label{DK}
\end{equation} 
which shows that $dK^{2}$ and $dL^{2}$ are always linearly independent
unless the coefficients of $dp_{r}$ and $dr$
both vanish simultaneously. From 
(\ref{INVR}) it follows  that $p_{r}$ vanishes at 
those $r$ that satisfy $\frac {\lambda^{2}}{a_{\theta}^{2}}=V(r)=(rf)^{-2}$, i.e. at $r_{\pm}$,
while for the  $k=1$ case the coefficient of $dr$ in 
(\ref{DK}) vanishes
at those $r$ that obey: $r_{+}=r_{-}=r_{min}=2$. Therefore, for  
$k=1$, the forms $dK^{2}$ and $dL^{2}$ fail to be 
linearly independent along the circular orbits supported by the 
potential $V(r)=(rf)^{-2}$. For the case of $k=0$, 
the forms $dK^{2}$ and $dL^{2}$ are always linearly independent
as long as $a^{2}_{\theta}>0$.\\

From this analysis 
we conclude\footnote{Here we ignore the special values of the parameters
that correspond to cases
 $\theta=\frac {\pi}{2}, p_{\theta}=0$  and 
$a^{2}_{\theta}=a^{2}_{\phi}$
as well the case of the circular orbits supported by the 
potential $V(r)=(rf)^{-2}.$}
of radial 
 that 
on the family of sets 
$\Gamma_{(m^{2}>0,~\lambda^{2}>0, ~a_{\theta}^{2}>0,~ a_{\phi})}$
always
$(dF_{1}\wedge dF_{2}\wedge dF_{3}\wedge dF_{4})\Big\vert _{(x,p)} \neq 0$. 
This in turn implies 
that these sets define 
a family of smooth 
four-dimensional  submanifolds  
of $T^{*}M$. For the case 
$k=0$
this family is
topologically 
$R\times R\times S^1\times S^1$
while for
 the case $k=1$ 
and as long as 
$\lambda^{2}>a_{\theta}^{2}>0$,
 then
$\Gamma_{(m^{2}>0,~\lambda^{2}>a_{\theta}^{2}>0,~ a_{\phi})}$ 
also defines a family of smooth 
four-dimensional  submanifolds  
topologically 
$R\times S^{1}\times S^1\times S^1$.\\
The conclusion that the
invariant submanifolds have topologies
$R\times R\times S^1\times S^1$
(case $k=0$)
and 
$R\times S^{1}\times S^1\times S^1$ (for $k=1$),
at a first sight
seems to contradict the
 celebrated property of integrable
Hamiltonian flows: their invariant  submanifolds are the Cartesian product of
 $n$-tori ( four-tori for our case).
 However, this property holds 
 whenever 
 the invariant submanifolds  
 are compact and connected
 (for a proof of this property see for instance 
 ref.\cite{1}, page 272).
 The motion on the $(t,p_{t})$ and $(r,p_{r})$ 
 planes for the $k=0$ case, fails to be bounded and this property is reflected in the topologies of the invariant
  submanifolds.
 In fact in refs.\cite{Shm1},\cite{OP1},\cite{OP2} they have also
 found  
 $R\times S^{1}\times S^1\times S^1$
topologies\footnote{Our thanks to Olivier Sarbach for discussing this point with us.}
 for the invariant submanifolds of  relativistic Hamiltonian integrable systems.\\

To complete the analysis of invariant submanifolds we now
 consider the family  
of sets
$$
\Gamma_{(m^{2}>0, ~\lambda^{2}>0, ~a_{\theta}^{2}=a_{\phi}=0)}=\{(x,p)\in T^{*}_{x}M, \ F_{1}(x,p)=2H(x,p)=-m^{2},
$$
\begin{equation}
F_{2}(x,p)=F_{3}(x,p)=0,\quad F_{4}(x,p)=\lambda^{2}>0 \}.
\label{INV}
\end{equation}
and at first we show that 
these sets are non empty.
For this, we employ 
the spherical gauge
and restrict our  attention
to points on the bundle coordinatized according to: 
$(x,p)=(t,r,\theta, \phi, p_{t}, p_{r}, 0,0)$.
Clearly for such points
$\hat \xi_{(i)}(x, p)=<p,\xi_{(i)}(x)>=0$ for all $ i=1,2,3$
(but note that in general for such points $\hat \eta_{(i)}(x,p)=<p,\eta_{(i)}(x)>\neq 0$ ).\\
An alternative
way to implement
$F_{2}(x,p)=F_{3}(x,p)=0$
is to employ the Cartesian gauge and 
restrict our attention to the points:
$(x,p)=(t,x,y,z, p_{t}, p_{x}=Ax, p_{y}=Ay, p_{z}=Az)$, where $A$ is an arbitrary but smooth function 
of $(x,p)$. Clearly, 
$F_{2}(x,p)=F_{3}(x,p)=0$ but also notice that $dF_{2}(x,p)=dF_{3}(x,p)=0$ at any $(x,p)$  lying
on the levels $F_{2}(x,p)=F_{3}(x,p)=0$.\\
Since here $m^{2}>0$ and $\lambda^{2}>0$,
 the levels of 
$F_{1}(x,p)=2H(x,p)=-m^{2}<0$ are those 
described by 
$(\ref{Re1})$
while the levels  of 
$K^{2}(x,p)=\lambda^{2}>0$ are topologically lines
satisfying:
$$
p_{r}^{2}-\lambda^{2}f^{2}(r)=0,\quad \lambda^{2}>0
$$
Moreover, for 
$a^{2}_{\theta}=a_{\phi}=0$, 
$(\ref{Cir1})$ implies that $p_{\theta}=0$ and thus $\frac {d\theta}{d\tau}=0$,
implying $\theta$ is constant, taken without loss of generality to 
have the equatorial value.\\
We now consider the forms
$(dH, dF_{2}, dF_{3}, dK^{2})$ 
evaluated on 
$\Gamma_{(m^{2}>0 ~\lambda^{2}>0, ~a_{\theta}^{2}=a_{\phi}=0)}$.
Since on these sets  
$dF_{2}=dF_{3}=0$ and
only $(dF_{1}\wedge dF_{4})\Big\vert _{(x,p)}\neq0$ is non vanishing 
we interpret this as implying that 
 the invariant submanifolds 
$\Gamma_{(m^{2}>0 ~\lambda^{2}>0, ~a_{\theta}^{2}=a_{\phi}=0)}$
are topologically $R\times R$, although we do not have
a rigorous mathematical proof of this claim.
As we have discussed in the introduction, the
topology of such singular invariant submanifolds is 
 a subtle problem (see refs. \cite{FO1},\cite{PE1})
 and at this point  more work is needed.\\

The family of the sets
$$
\Gamma_{(m^{2}>0, ~\lambda^{2}=0, ~a_{\theta}^{2}>0,~a_{\phi}\in R)}=\{(x,p)\in T^{*}_{x}M, \ 
F_{1}(x,p)=2H(x,p)=-m^{2},
$$
\begin{equation}
F_{2}(x,p)=a^{2}_{\theta}, F_{3}(x,p)=a_{\phi},\quad F_{4}(x,p)=\lambda^{2}=0 \}. 
\label{INV2}
\end{equation}
seems to be empty.
This follows by noting that
$F_{4}(x,p)=\lambda^{2}=0 \Longrightarrow K^{2}(x,p)+kL^{2}(x,p)=0$
and thus by appealing to 
(\ref{SF3}) we conclude that
$$p_{r}^{2}=-\frac {a^{2}_{\theta}}{r^{2}}$$
which leads to  a contradiction unless $a^{2}_{\theta}=0$.\\

Interestingly however,
the set $\Gamma_{(m^{2}>0, ~\lambda^{2}=a_{\theta}^{2}=a_{\phi}=0)}$ is 
non empty.
This can be seen by 
restrict attention on the points  $(x,p)$ coordinatized
according to
$(x,p)=(t,x,y,z,p_{t}, 0,0,0)$.
Clearly at such points
$\hat \xi_{(i)}(x, p)=\hat \eta_{(i)}(x,p)=0$ for all $ i=1,2,3$ and in this case
the only non trivial level sets are those described by
$(\ref{Re1})$. 
Evaluating again
$(dH, dF_{2}, dF_{3}, dK^{2})$ 
on 
$\Gamma_{(m, ~\lambda^{2}=a_{\theta}^{2}=a_{\phi}=0)}$
then clearly the only non vanishing differential is $dF_{1}=2dH$
implying that the invariant submanifolds are one dimensional
described by the family of lines shown in 
(\ref{InvT}) in the limit of vanishing $\lambda$. \\

\section{Summary and Discussion}
In this work we have analyzed
the structure of the timelike component of the geodesic flow
for the family of spatially flat and spatially closed FRW spacetimes
and in this section we discuss 
the benefits of this analysis.\\
In order to do so, 
it is instructive at first  
to discuss
the connection
between a geodesic flow and
 the notion of a single geodesic.
This connection is best illustrated
by recalling the analysis of the Kepler problem in Classical Mechanics
(see for example ref.\cite{3}).
For a single particle moving  in the attractive
Newtonian spherical potential, the possible orbits are well known.
Due to the symmetries, one can always choose
 a plane that  contains the entire orbit
and in that way 
all details of the orbits are are easily determined.
 However the situation becomes more complex
(and also more interesting), when more than one test 
particles  are involved.
In that case,
 one considers Hamiltonian methods and constructs the
 corresponding
Hamiltonian flow and associated invariant subspaces 
along the lines discussed for instance in section $10$ of ref. \cite{3}.
Once the structure of these invariant submanifolds are known,
they provide information
regarding the structure of the trajectory through 
any chosen point on the phase space.\\

 Similar situation occurs for the case of causal geodesics on an FRW spacetime.
The large number of the Killing fields admitted by this background makes
the analysis of the behavior of (a single) causal geodesic a trivial problem.
 Using these Killing fields,
one readjusts the coordinate gauge so that 
for  any chosen initial condition,
the motion takes place either on the $(t, r)$-plane or
 along the direction orthogonal to the family 
 of the hypersurfaces  which are  homogeneous and isotropic
 (see for instance  discussion in ref.
 \cite{Wald}, page 103 or consult 
\cite{COS1},\cite{COS2}).
 The first family of timelike geodesics
  corresponds to the choices $m^{2} >0$, $\lambda^{2}>0$,
 $a^{2}_{\theta}=a_{\phi}=0$ 
 while the second family to the choice of
 constants: $m^{2}>0$, $\lambda^{2}=a^{2}_{\theta}=a_{\phi}=0$
 (see eqs.(\ref{FIRI}). 
 
 But, like for the case of
  the Kepler problem, the
 situation  becomes  more involved if we assign over a hypersurface which is homogeneous
  and isotropic
 a distribution of initial positions and of timelike future pointing four momenta (velocities)
 as for example occurs in problems dealing with relativistic  kinetic theory.
For this case, one cannot any longer readjust
  the coordinate gauge so that
 all particles move on the same $(t,r)$ plane or comoving with 
 the expansion of the universe.
 A framework to address these problems is provided
 by the  structure 
 of the timelike component of the geodesic flow
 carried out in this work. If for instance, the initial 
 distribution has its support
  in the invariant submanifold
described by $\Gamma_{(m^{2}>0,~\lambda^{2}>0,~a_{\theta}^{2}>0,~a_{\phi})}$ 
 then we have an understanding
  of the behavior of the phase space trajectories.
  As we have shown in this work,  
 they will  lie entirely within
 $\Gamma_{(m^{2}>0,~\lambda^{2}>0,~a_{\theta}^{2}>0,~a_{\phi})}$ 
 and the topology of that submanifold provides information regarding the 
 global behavior of the these phase space trajectories. By projecting them on the base $(M,g)$
 we obtain information regarding
 the structure of the corresponding timelike geodesics
 and thus for a collisionless gas, its future evolution.
 In conclusion, the structure of the geodesic flow offers
 insights not only to the behavior
 of a single geodesic, but rather
 to a family of such geodesics. \\
 The results of this paper 
can be used to construct the most general 
solutions of the collisionless Louiville equation on an FRW background
by taking  advantage 
of the generating function
constructed by the complete integral of the 
 Hamilton-Jacobi equation on an FRW background.
 Via this function, we introduce 
suitable coordinates in the cotangent bundle
and proceed
along the same lines as in 
\cite{OP1}, \cite{OP2}
extending also  the work carried out in \cite{O4}.\\
 
 Although the present work addressed some 
 properties of the timelike component of the geodesic flow
 mainly for the spatially flat and spatially closed universe, it would be of interest
 to address  the properties of the geodesic flow
 on a spatially hyperbolic universe. Moreover, it would of interest  to address
 the structure of the null component of the geodesic flow on an FRW background
 since as  we have
 noticed at various stages of this work, the nullity condition introduces some
 new challenges that are worth to be considered.
 We hope to discuss these issues in the future.\\
 \\
Finally, this work shows that the geodesic flow on an  FRW background is a soluble model
and this is welcomed. It offers the possibility to understand subtle issues
regarding the dynamics of completely integrable relativistic Hamiltonians
such as the foliation of the relativistic  phase space by invariant submanifolds, singularities in these foliations,
etc. As we have already mentioned, it would be of interest
 to understand the structure
of the invariant submanifolds over points where the integrals in involution fail to be independent.
The results of the paper may offer some clues toward this direction due to the
presence of the symmetries, but definitely more work is needed in this direction.

The present work was focused on the geodesic flow on a highly symmetric background spacetime, namely 
the family of spatially flat homogenous and isotropic spacetimes. It would be very interesting to
investigate the  Liouville-Arnold type of integrability for the geodesic flow
on less symmetric backgrounds such as the family of Bianchi models.
In that regard, the authors\footnote{Our thanks to an anonymous referee for bringing to our attention refs
\cite{B1},\cite{B2}.}  of refs \cite{B1},\cite{B2} introduce a particular family of spacetimes
referred as 
St\"ackel spacetimes having the property that the Hamilton-Jacobi equation is separable.
In ref.\cite{B2}, they define a particular family of spatially homogeneous (but non spatialy isotropic) St\"ackel spacetimes
that includes as a special case the family of Bianchi cosmologies
and have proven separability of Hamilton-Jacobi for geodesic motion.
 It would be interesting
 to examine whether the resulting integrals are globally well defined
 and study the structure of the invariant by the flow submanifolds.
It is our hope that the present work, combined with that in \cite{B2}, the analysis in  
\cite{Shm1} and the studies of the geodesic flow on a Schwarzschild black hole \cite{OP1},
\cite{OP2}  may act as a further stimulus for a systematic study of the fascinating subject of relativistic geodesic flows.

\section*{Acknowledgments}
Our pleasure to thank Olivier Sarbach and Paola Rioseco for many stimulating discussions regarding the issues raised in this work.
The research of F.A. and T.Z was supported  by CIC grant from the Universidad Michoacana, 
and CONACYT Network Project 280908 Agujeros Negros y Ondas 
Gravitatorias. J.F.S thanks CONACyT for a predoctoral  fellowship.\\
 Finally our thanks to an anonymous referee for his constructive comments and criticism that improved the overall presentation of the the paper.
\\\

\end{document}